\documentstyle{aipproc}
\begin{document}
\title{Physics Opportunities Above the \\ Greisen-Zatsepin-Kuzmin Cutoff: \\ 
Lorentz Symmetry Violation \\ at the Planck Scale}

\author{Luis Gonzalez-Mestres$^{*,\dagger}$}
\address{$^*$Laboratoire
de Physique Corpusculaire, Coll\`ege de France, 75231 Paris 
Cedex 05, France\\
$^{\dagger}$L.A.P.P., CNRS-IN2P3, 
B.P. 110, 74941 Annecy-le-Vieux Cedex, France}

%\lefthead{LEFT head}
%\righthead{RIGHT head}
\maketitle

\begin{abstract}
Special relativity has been tested at low energy with great accuracy, but
these results cannot be extrapolated to very high-energy phenomena:
this new domain of 
physics may actually provide the key to the, yet unsettled, 
question of the ether and the absolute rest frame.
Introducing a critical distance scale, $a$ ,
below $10^{-25}~ cm$ (the wavelength
scale of the highest-energy observed cosmic rays) allows to consider models,
compatible with standard tests of special relativity, where a small
violation of Lorentz symmetry ($a$ can, for instance, be the Planck length)
leads to a deformed relativistic kinematics (DRK)
producing dramatic effects on the properties 
of very high-energy cosmic rays.
For instance, the Greisen-Zatsepin-Kuzmin (GZK)
cutoff 
does no longer apply and particles which are unstable at low energy
(neutron, some hadronic resonances
like the $\Delta ^{++}$, possibly several nuclei...)
become stable at very high energy.
In these models, an absolute local rest frame exists (the {\bf vacuum rest
frame}, VRF) and special relativity is a low-momentum limit. 
We discuss the possible effects of Lorentz symmetry violation (LSV)
on kinematics and dynamics, 
as well as the cosmic-ray energy
range (well below the energy scale associated to the fundamental length)
and experiments (on earth and from space) where they could be detected.
\end{abstract}

\pagenumbering{arabic}

\section*{STATUS OF THE RELATIVITY PRINCIPLE}

H. Poincar\'e was the first author to consistently
formulate the relativity principle on the grounds of experiment, stating in
1895 \cite{Poincare95}:

{\it "Experiment has provided numerous facts justifying the following
generalization: absolute motion of matter, or, to be more precise, the
relative motion of weighable matter and ether, cannot be disclosed. All that
can be done is to reveal the motion of weighable matter with respect to
weighable matter"}

The deep meaning of this law of Nature was further
emphasized by the same author when he wrote in 1901 \cite{Poincare01},
in connexion with Lorentz contraction:

{\it "Such a strange property seems to be a real coup de pouce presented by
Nature itself, for avoiding the disclosure of absolute motion...
I consider quite probable that optical phenomena depend only on
the relative motion of the material bodies present, of the sources of 
light and optical instruments, and this dependence is not accurate...
but rigorous.
This principle will be confirmed with increasing precision,
as measurements
become more and more accurate"}

The role of H. Poincar\'e in building relativity,
and the relevance of his thought, have
often been emphasized \cite{Logunov95,Fey}.
In his
June 1905 paper \cite{Poincare05}, published before Einsteins's article
\cite{Einstein05} arrived (on June 30) to the editor,
he explicitly wrote the relativistic transformation law for the
charge density and velocity of motion and applied to gravity
the "Lorentz group" (that he introduced), 
assumed to hold for "forces of whatever origin".
From this, he inferred that "gravitational waves" propagate at the speed
of light. 
However, his priority is sometimes
denied \cite{Miller96,Paty96} on the grounds that {\it "Einstein essentially
announced the failure of all ether-drift experiments past and future as a
foregone conclusion, 
contrary to Poincar\'e's empirical bias"} \cite{Miller96},
that Poincar\'e did never {\it "disavow the ether"} \cite{Miller96} or that
{\it "Poincar\'e never challenges... the absolute time of newtonian
mechanics... the ether is not only the absolute space of mechanics... but a
dynamical entity"} \cite{Paty96}. It is implicitly assumed that A. Einstein
was right in 1905 when {\it "reducing ether to the absolute space of
mechanics"} 
\cite{Paty96} and that H. Poincar\'e was wrong because {\it "the
ether fits quite 
nicely into Poincar\'e's view of physical reality: the ether
is real..."} \cite{Miller96}. A basic physics issue (whether ether  
and an absolute rest frame exist or
not), perhaps not definitely settled, underlies the debate on priority.
Actually, modern particle physics has
brought 
back the concept of a non-empty vacuum where free particles propagate:
without such an "ether" where fields can condense, the standard model
of electroweak interactions could not be written and quark confinement could
not be understood. Modern cosmology is not incompatible with
an "absolute local frame" (the {\bf vacuum rest frame}, VRF)
close to that suggested by the study of cosmic
microwave background radiation. If "ether" and the VRF actually exist, 
the relativity principle (the impossibility to disclose absolute motion)
will become
a symmetry, a concept whose paternity was attributed to
H. Poincar\'e by R.P. Feynman \cite{Feynman}:

{\it "Precisely Poincar\'e proposed investigating what could 
be done with the
equations without altering their form. It was precisely his idea to pay
attention to the symmetry properties of the laws of physics"}

As symmetries in particle physics are in general violated at some scale,
Lorentz symmetry may be broken and an absolute local rest frame may
be detectable through experiments performed beyond the relevant scale.
It may even happen that Lorentz symmetry be just an infrared attractor.
Poincar\'e's special relativity (a symmetry applying to
physical processes) could live with this situation, in which case the
relativity principle would refer to the impossibility to disclose absolute
motion {\it through low-energy experiments}. But Einstein's
approach, such as it was formulated in 1905 (an 
absolute geometry of space-time
that matter cannot escape), could not survive. 
We discuss here two issues: a) the scale
where we may expect Lorentz 
symmetry to be violated; b) the physical phenomena
and experiments 
potentially able to uncover Lorentz symmetry violation (LSV).
Previous papers on the subject are references \cite{Gon96} to \cite{Gon9711}
and references therein.

\section*{RELATIVITY AS A LOW-ENERGY LIMIT}

Low-energy tests of special relativity have confirmed its validity to an
extremely good accuracy
\cite{Lam,Hills}, in impressive agreement with Poincar\'e's 1901 conjecture. 
But the situation at very high energy remains more
uncertain (see \cite{Gon96} to \cite{Gon9711}): high-energy physics 
corresponds to a domain never covered by the experiments 
that motivated special relativity a century ago. Figures can 
change by more than 20 orders of magnitude between the highest oberved 
cosmic-ray energies and the scale explored by the above mentioned tests
of Lorentz symmetry.
If LSV follows
a $E^2$ law
($E$ = energy), similar to the effective gravitational coupling, it can
be of order 1 at $E~\approx ~10^{21}~eV$ and $\approx ~10^{-26}$ at $E~
\approx ~100~MeV$ 
(corresponding to the highest momentum scale involved in
nuclear magnetic resonance experiments), in which case it will escape all
existing low-energy bounds. If LSV is $\approx 1$ at Planck scale ($E~
\approx ~10^{28}~eV$), and following a similar law, it will be $\approx
~10^{-40}$ at $E~\approx ~100~MeV$ . Our suggestion is not in contradiction
with Einstein's thought after he had developed general
relativity. In 1921 , A. Einstein wrote \cite{Ein21}:

{\it "The interpretation of geometry advocated here cannot be 
directly applied
to submolecular spaces... it might turn out that such an extrapolation is
just as incorrect as an extension of the concept of temperature to particles
of a solid of molecular dimensions"}.

It is in itself remarkable that special relativity holds at the
attained accelerator energies, thus confirming Poincar\'e's conjecture 
far beyond expectations. But there is no fundamental reason for this 
dazzling success to persist
above Planck scale.
A typical (and natural) 
example of models violating Lorentz symmetry at very short distance
is provided by models where an absolute local rest frame exists and
non-locality in space is introduced through a
fundamental length scale $a$ \cite{Gon9703}. Such models lead in the VRF 
to a deformed
relativistic kinematics of the form \cite {Gon9703,Gon9710}:
\equation
E~=~~(2\pi )^{-1}~h~c~a^{-1}~e~(k~a)
\endequation
\noindent
A
where $h$ is the Planck constant, 
$c$ the speed of light, $k$ the wave vector
and
$[e~(k~a)]^2$ is a convex
function of $(k~a)^2$ obtained from vacuum dynamics.
Expanding equation (1) for $k~a~\ll ~1$ , we can write:
\equation
e~(k~a) ~~ \simeq ~~[(k~a)^2~-~\alpha ~(k~a)^4 
~+~(2\pi ~a)^2~h^{-2}~m^2~c^2]^{1/2}
\endequation
\noindent
$m$ being the mass, $\alpha $ a model-dependent positive constant
$\approx 0.1~-~0.01$ for
full-strength LSV at
momentum scale $p~\approx ~ a^{-1}~h~$,
and in terms of momentum $p$~:
\equation
E ~~ \simeq ~~ p~c~+~m^2~c^3~(2~p)^{-1} 
~-~p~c~\alpha ~(k~a)^2/2~~~~~
\endequation

The "deformation" $\Delta ~E~=~-~p~c~\alpha ~(k~a)^2/2$ in the right-hand
side of (3) implies a Lorentz symmetry violation in the ratio $E~p^{-1}$
varying like $\Gamma ~(k)~\simeq ~\Gamma _0~k^2$ where $\Gamma _0~
~=~-~\alpha ~a^2/2$ . If $c$ and $\alpha $ are universal parameters for all
particles, LSV does not lead to the spontaneous decays predicted in
\cite{Col}: the existence of very high-energy cosmic rays
cannot be regarded as an evidence against LSV. With the deformed 
relativistic
kinematics (DRK) defined by (1)-(3), Lorentz symmetry remains valid
in the limit $k~\rightarrow ~0$, contrary to the standard $TH\epsilon \mu $
model \cite{Will}. The above non-locality may actually be an 
approximation to
an underlying dynamics involving superluminal particles
\cite{Gon96,Gon9710,Gon9701,Gon9705,Gon9709}, 
just as electromagnetism looks nonlocal
in the potential approximation to lattice dynamics in solid-state physics:
it would then correspond to the limit $c~c_i^{-1}~\rightarrow~0$
where $c_i$ is the superluminal critical speed.

As recently pointed out \cite{Gon9704}, equation (1) 
is a fundamental property of old
scenarios (f.i. \cite{Red}) breaking local Lorentz invariance (LLI).
An ansatz based on an isotropic, continuous modification of the Bravais
lattice dynamics is \cite{Gon9703}:
\equation
e~(k~a)~~=~~[4~sin^2~(ka/2)~+~(2\pi ~a)^2~h^{-2}~m^2~c^2]^{1/2}
\endequation
and simple extensions of the ansatz by R\'edei 
\cite{Red}
lead \cite{Gon9704} to expressions like:
\begin{eqnarray}
e~(k~a)~~=~~[10~+~30~(k~a)^{-2}~cos~(k~a)~-~~~~~~~~~~~~~~~ \nonumber \\
~~~~~~~~~~~~~~~~~30~(k~a)^{-3}~sin~(k~a)
~+~(2\pi ~a)^2~h^{-2}~m^2~c^2]^{1/2}
\end{eqnarray}

In any case, we expect observable kinematical effects when the term
$\alpha (ka)^3/2$ becomes as large as the term
$2~\pi ^2~h^{-2}~k^{-1}~a~m^2~c^2$ .
This happens at:
\equation
E~~\simeq (2\pi )^{-1}~h~c~k~~\approx ~~ E_{trans}~~\approx~~ 
\alpha ^{-1/4}~
(h~c~a^{-1}/2\pi)^{1/2}~(m~c^2)^{1/2}
\endequation

Thus, contrary to conventional estimates
of LLI breaking predictions \cite{Doqui}
where the modification of
relativistic kinematics is ignored,
observable effects will be produced at wavelength scales well above
the fundamental length.
For a nucleon, taking $a~\approx ~10^{-33}~cm$ and $\alpha ~\approx 0.1$,
this corresponds to $E~\approx 10^{19}~eV$~, below the highest
energies at which cosmic rays have been observed. 
With full-strength LSV, for a proton
at $E~\approx ~ 10^{20}~eV$ and with the above value of $a$ , we get:
\equation
\alpha ~(k~a)^2/2~~\approx ~~10^{-18} 
~~\gg ~~2~\pi ^2~h^{-2}~k^{-2}~m^2~c^2~~
\approx ~~10^{-22}
\endequation
and, although $\alpha (ka)^3/2$ is very small as compared to the
value of $e~(k~a)$ , the term $2~\pi ^2~h^{-2}~k^{-1}~a~m^2~c^2$ represents
an even smaller fraction. 
Although relativity reflects to a very good approximation
the reality of physics at large distance scales and can be
considered as its low-energy limit, no existing
experimental result proves that it applies with the same accuracy to
high-energy cosmic rays. 

Are $c$ and $\alpha $ universal? This may be
the case for all "elementary" particles, i.e.
quarks, leptons, gauge bosons...,
but the situation is less obvious for hadrons, nuclei and heavier objects.
From a naive soliton model \cite{Gon9703}, we inferred that: a) $c$ is
expected to be universal up to very small corrections ($\sim 10^{-40}$)
escaping all existing bounds; b)
a possible
approximate rule can be to take $\alpha $ universal for leptons, 
gauge bosons
and light hadrons (pions, nucleons...) and assume a $\alpha \propto m^{-2}$
law for nuclei and heavier objects, the nucleon mass setting the scale.

\section*{RELEVANCE OF COSMIC-RAY EXPERIMENTS}

If Lorentz symmetry is broken at Planck scale or at some other
fundamental scale, and assuming that the earth moves slowly with
respect to the VRF,
the effects of LSV may be observable well below this
energy and produce detectable phenomena at the highest
observed cosmic-ray energies. This is due to DRK
\cite{Gon9703,Gon9710,Gon9708,Gon9706}: at energies above
$E_{trans}$ ,
the deformation $\Delta ~E$
dominates over
the mass term $m^2~c^3~(2~p)^{-1}$ in (3) and modifies all
kinematical balances. Because of the negative value of $\Delta ~E$ , 
it costs
more and more energy, as energy increases above $E_{trans}$,
to split the incoming logitudinal momentum.
The parton model (in any version), as well as standard
formulae for Lorentz contraction and time dilation, are also 
expected to fail
above this energy \cite{Gon9710,Gon9708} which corresponds to $E
~\approx~10^{20}~eV$
for $m$ = proton mass and
$\alpha ~a^2~\approx ~10^{-72}~cm^2$ (f.i. $\alpha
~\approx ~10^{-6}$ and $a$ = Planck
length), and to $E~\approx ~10^{18}~eV$ for
$m$ = pion mass and $\alpha ~a^2~\approx ~10^{-67}~cm^2$
(f.i. $\alpha ~\approx ~0.1$ and $a$ = Planck length).
In particular, the following effects are predicted:

a) For $\alpha ~a^2~>~10^{-72}~cm^2$ , and
assuming a universal value of $\alpha $ ,
the Greisen-Zatespin-Kuzmin (GZK)
cutoff \cite{Greisen,Zatsepin} is suppressed 
\cite{Gon9703,Gon9708,Gon9704,Gon9706} 
for the particles under
consideration and ultra-high energy cosmic rays (e.g. protons)
produced anywhere in the presently observable Universe
can reach the earth without losing their energy in collisions with the 
cosmic microwave background radiation.

b) With the same hypothesis,
unstable particles with at
least two stable particles in the final states
of all their decay channels become stable at very
high energy \cite{Gon9703,Gon9704}. Above $E_{trans}$, the lifetimes of all
unstable particles (e.g. the $\pi ^0$ in
cascades) become much longer than predicted
by relativistic kinematics \cite{Gon9703,Gon9704,Gon9706}.

c) In astrophysical processes at very
high energy,
similar mechanisms can inhibit radiation under
external forces, GZK-like cutoffs, decays,
photodisintegration of nuclei, momentum loss through
collisions, production of lower-energy secondaries...
potentially contributing to solve all basic problems
raised by the highest-energy cosmic rays. Therefore, calculations of 
astrophysical processes at very high energy cannot ignore the possibility 
that Lorentz symmetry be violated \cite{Gon9706}.

d) With the same hypothesis, the allowed final-state
phase space of two-body collisions is modified and
can lead to a sharp fall of cross-sections
for incoming cosmic ray energies above
$E_{lim} ~\approx ~(2~\pi )^{-2/3}~(E_T~a^{-2}~ 
\alpha ^{-1}~h^2~c^2)^{1/3}$,
where $E_T$ is the energy of the target \cite{Gon9707}. 
As a consequence, and with the
previous figures for the parameters of LSV, above some
energy $E_{lim}$ between 10$^{22}$ and $10^{24}$ $eV$ a cosmic
ray will not deposit most of its energy in the atmosphere
and can possibly fake an exotic event with much less energy.

e) Effects a) to d) are obtained using only DRK. If dynamical
anomalies are added (failure, at very small distance scales,
of the parton model and of the
standard Lorentz formulae for length and time \cite{Gon9710,Gon9708}...), 
we can expect
much stronger effects in the cascade development profiles
of cosmic-ray events.

f) Cosmic superluminal particles would produce atypical events
with very small total momentum, isotropic or involving several
jets \cite{Gon96,Gon9710,Gon9705,Gon9709}.

In what follows, we discuss in more detail the implications of these effects.

\section*{THE GZK CUTOFF DOES NO LONGER APPLY}

For $\alpha ~a^2~>~10^{-72}~cm^2$, with a universal value 
of $\alpha $ , a $E~\approx ~10^{20}~ eV$ proton interacting with a cosmic 
microwave background photon would be sensitive 
to DRK effects \cite{Gon9703}. 
After having absorbed a $10^{-3}~ eV$ 
photon moving in the opposite direction,
the proton gets an extra $10^{-3}~ eV$ energy, whereas its momentum is
lowered by $10^{-3}~ eV/c$ . In the conventional scenario with exact Lorentz
invariance, this is enough to allow the excited proton to decay into a
proton or a neutron
plus a pion, losing an important part of its energy. 
The small increase in the $E/p$ ratio is enough to generate, in the final 
state, the increase of the nucleon mass term 
$m^2~(2~p)^{-1}$ (as momentum gets lower)
as well as the pion mass term and the transverse energy of both particles.
However, it
can be checked \cite{Gon9703} 
that in our scenario with LSV such
a reaction is strictly forbidden, as the $\approx 2.10^{-23}$ increase 
of the $E/p$ ratio cannot provide the energy required, 
due to the deformation term $\Delta ~E$ , by the splitting of
the incoming momentum.  
Elastic $p~+~\gamma $ scattering
is permitted, but allows the
proton to release only a small amount of its energy.
Similar or more stringent bounds exist
for channels involving lepton production. Obvious phase
space limitations will also lower the collision rate, as compared to standard
calculations using exact Lorentz invariance which predict photoproduction
of real pions at such cosmic proton energies. The effect is  
strong enough
to invalidate the GZK cutoff
and explain the
existence of the highest-energy cosmic-ray events. It will become more
important at higher energies, as we get closer to the $a^{-1}$ wavelength
scale. If $\alpha $ is not universal, the particle with the highest
value of $\alpha $ can always reach the earth, and other particles can below 
some energy above the GZK cutoff. 
Assuming exact universality of $c$ , the situation for 
nuclei will crucially depend on the precise values of
$\alpha $ for each nucleus
and on the energy range. Models where $c$ is not exactly universal are not
ruled out, as the results from \cite{Col} do not apply \cite{Gon9701} , 
and deserve 
cosideration \cite{Gon9706}. 
Our result is limited by the history of the Universe, as
cosmic rays coming from distances closer and closer
to horizon
size will be older and older and, at early times, will have been confronted
to rather different scenarios. But it is clear that DRK 
allows much older ultra-high energy cosmic rays, generated at much more
remote sources, to reach the earth nowadays.

A previous attempt to explain the experimental absence of the predicted
GZK cutoff by
Lorentz symmetry violation at high energy \cite{Kirzhnits}
proposed an ansatz replacing relativistic kinematics by the relation:
\equation
E~~~=~~~m~~h~(p^2~E^{-2})
\endequation
where the positive function $h$ tends to $(1~-~p^2~E^{-2})^{-1/2}$ in the 
relativistic limit. These authors
considered an expansion in powers of $\gamma ^4$ , where
$\gamma ~=~(1~-~v^2c^{-2})^{-1/2}$~, $v$ is the speed of the particle
and the coefficient of the linear term
in $\gamma ^4$ had to be arbitrarily
tuned to $\approx 10^{-44}$ in order to produce
an effect of order 1 for a $10^{20}~eV$ proton (leading to a
potentially divergent expansion at higher energies). 
No such problems are encountered
in our approach, where the required orders of magnitude come out 
naturally in terms of small perturbations. 

\vskip 5cm

\section*{LIFETIMES AT VERY HIGH ENERGY}

In standard relativity, we can compute the lifetime of any unstable particle
in its rest frame and, with the help of a Lorentz transformation, obtain
the Lorentz-dilated lifetime for a particle moving at finite speed.
The same procedure had been followed in previous estimates 
of the predictions
of LLI breaking
\cite{Doqui}
for the decay of high-energy particles.
This is no longer possible with the 
kinematics defined by (1), which explicitly incorporates LSV.
Instead, two results are obtained \cite{Gon9703,Gon9704}:

i) Assuming universal values of $c$ and $\alpha $ , 
unstable particles with at least two massive particles in the final state
of all their decay channels become stable at very high energy,
as a consequence of the effect of LSV through (1).
A typical order of magnitude for the energy $E^{st}$
at which such a phenomenon occurs
is:
\equation
E^{st}~~\approx ~~c^{3/2}~
h^{1/2}~(a~m_2)^{-1/2}~(m^2~-~m_1^2~-~m_2^2)^{1/2}
\endequation
where: a) $m$ is the mass of the decaying particle; b)
we select the two
heaviest particles of the final product of each decay channel, and
$m_2$ is the mass of the lightest particle
in this list; c) $m_1$ is the mass of the 
heaviest particle produced together
with that of mass $m_2$ .
With $a~\approx ~10^{-33}~cm$ and $\alpha ~\approx ~0.1$ 
for all particles, {\bf the neutron
would become stable} above $E~\approx ~10^{20}~eV$ .
{\bf Some hadronic resonances} (e.g. the $\Delta ^{++}$ , whose
decay
product contains a proton and a positron)
{\bf would become stable} above
$E~\approx ~10^{21}~eV$ . Similar considerations may apply to some 
supersymmetric particles. 
Most of these objects will
decay before they can be accelerated to such energies, but
they may result of a collision at very high energy or of
the decay of a
superluminal particle \cite{Gon96},
The study of very high-energy cosmic rays can thus
reveal as stable particles objects which would be unstable if produced at
accelerators.
If one of the light neutrinos
($\nu _e$ , $\nu _{\mu }$) has a mass in the $\approx ~10~eV$
range, {\bf the muon would
become stable} at energies above $\approx 10^{22}~eV$~. Weak neutrino mixing
may restore muon decay, but with very long lifetime. {\bf
Similar considerations
apply to the $\tau $ lepton},
which would become stable above
$E~\approx 10^{22}~eV$ if 
the mass of the $\nu _{\tau }$ is $\approx ~100~eV$
but, again, a decay with very long lifetime can be restored by
neutrino oscillations. For nuclei, the situation will depend on the details of
DRK (basically, the value of $\alpha $ for each nucleus)
and deserves further investigation using more precise theoretical models.

ii) With the same hypothesis as i), all 
unstable particles live longer than naively expected with exact
Lorentz invariance and, at high enough energy,
the effect becomes much stronger than previously estimated
\cite{Doqui} ignoring the effects of DRK.
At energies well below the stability region, partial decay rates are already
modified by large factors leading to observable effects.
Irrespectively of whether
$m_2$ vanishes or not, the phenomenon
occurs 
above $E~\approx ~c^{3/2}~h^{1/2}~(m^2~-~m_1^2)^{1/4}~a^{-1/2}$
($\approx 10^{18}~eV$ for $\pi ^+~\rightarrow ~\mu ^+~+~\nu _{\mu }$~, if
$a~\approx ~10^{-33}~cm$ and $\alpha ~\approx ~0.1$). 
The effect has a sudden, sharp rise, since a fourth
power of the energy is involved in the calculation. In the LSV scenario, 
partial branching ratios
become energy-dependent.

\vskip 5cm

\section*{FINAL-STATE PHASE SPACE}

No special constraint seems to arise from (2) if, in the VRF, two
particles with equal,
opposite momenta of modulus $p$ with $\alpha ~(k~a)^2~
\ll ~1$ collide to produce a multiparticle final state.
But, as a consequence
of
LSV,
the situation becomes fundamentally different
at very high energy if
one of the incoming
particles is close to rest with respect to the VRF 
where formulae (1)~-~(3) apply.
Assume a very high-energy particle (particle 1)
with momentum ${\vec {\bf p}}$ , impinging on a particle
at rest (particle 2). We take
both particles to have mass $m$ , and $p~\gg ~mc$~.
In relativistic kinematics, 
we would have elastic final states where particle
1 has, with respect to the direction of ${\vec {\bf p}}$ ,
longitudinal momentum $p_{1,L}~\gg ~mc$ and particle 2 has
longitudinal momentum
$p_{2,L}~\gg ~mc$  with $p_{1,L}~+~p_{2,L}~=~p$ . A total transverse energy
$E_{\perp }~\simeq ~mc^2$ would be left for the outgoing particles.
But the balance is drastically modified by DRK 
if $\alpha ~(k~a)^2~p$ 
becomes of the same order
as $m~c$ or larger.
As
the energy increases, stronger and stronger limitations of the available
final-state phase space appear:
the final-state configuration $p_{1,L}~=~p~-~p_{2,L}~=~(1~-~\lambda )~p$
becomes kinematically forbidden for $\alpha ~(k~a)^2~p~>~2~m~c~\lambda ^{-1}
(1~-~\lambda )^{-1}/3$ . Above $p~\approx
(m~c~a^{-2}~h^2)^{1/3}$~,
"hard" interactions become severely
limited by
kinematical constraints.
Similarly, with the same initial state,
a multiperipheral final state configuration with $N$ particles ($N~>~2$)
of mass $m$ and longitudinal momenta $g^{i-1}~p'_L$
($i~=~1,...,N$ , $g~>
~1$),
where $p'_L~=~p~(g~-~1)~(g^{N}~-~1)^{-1}$~, $g^N~\gg ~1$
and $p'_L~\gg m~c$ , would have in
standard relativity an allowed total transverse energy
$E_{\perp }~(N~,~g)~\simeq ~
m~c^2~[1~-~m~c~(2~p'_L)^{-1}~(1~-~g^{-1})^{-1}]$ which is positive definite.
Again, using the new kinematics and the approximation (3), we find that
such a longitudinal final state configuration is forbidden for
values of the incoming momentum such that
$\alpha ~(k~a)^2~p~c~>~2~(3~g)^{-1}~(1~+~g~+~g^2)~E_{\perp }~(N~,~g)$~.
{\it Elastic, multiparticle and total cross sections will sharply fall at very 
high energy.}

For "soft" strong interactions, the approach were the two-body
total cross section is the less sensitive to final-state phase space
is, in principle, that based on dual resonance models and considering the
imaginary part of the elastic amplitude as being dominated by the shadow
of the production 
of pairs of very heavy resonances of masses $M_1$ and $M_2$
of order $\approx (p~m~c^3/2)^{1/2}$ in the direct channel
\cite{Aur1,Aur2}).
Even in this scenario, we find important limitations to the
allowed values of $M_1$ and $M_2$~, and to the two-resonance phase space,
when $\alpha ~(k~a)^2~p$ becomes $\approx m~c$ or larger.
Above $E~\approx (2\pi )^{-2/3}~(m~\alpha ^{-1}~a^{-2}~h^2~c^4)^{1/3}$~, 
nonlocal
effects play a crucial role and invalidate considerations based
on Lorentz invariance and local field theory used to derive the Froissart
bound \cite{Froissart}, which seems not to be violated but ceases
to be significant given the expected behaviour of total
cross sections which, at very high-energy,
seem to fall far below this bound.
An updated study of noncausal dispersion relations \cite{Blok1,Blok2},
incorporating
DRK and nonlocal dynamics, can
possibly lead to new bounds. If 
the target is not at rest in the VRF, but its
energy is small as 
compared to that of the incoming particle, its rest energy
must be replaced by the actual target energy $E_T$ in the above discussions.
The absence of GZK cutoff is a particular application of this analysis,
which has a much general validity.

\section*{EXPERIMENTAL IMPLICATIONS}

With DRK, a small violation of the universality of $c$ would not
necessarily produce the Cherenkov effect in vacuum
considered by Coleman and Glashow \cite{Col} for high-energy cosmic rays.
If $c$ and $\alpha $ are both universal and $\alpha $ is positive, 
all stable particles
remain stable when accelerated to ultra-high energies and can reach any energy
without spontaneously decaying or radiating in vacuum
(the case $\alpha ~<~0$ , not considered here, would lead to spontaneous 
decays and "Cherenkov" phenomena for particles at very high energy). 
If $c$ is universal but
$\alpha $ (positive) is not,
there will in any case be at least 
one stable particle at ultra-high energy (that with the highest positive value
of $\alpha $). If none of the two constants is 
universal, any scenario is {\it a priori} possible.
The mechanisms we just described compete with those considered
in \cite{Col} and tend to compensate their effect:
therefore, the bounds obtained by these authors do not apply to our
ansatz where small violations of the universality of $c$ can be compensated 
by the deformation term \cite{Gon9701,Gon9706}. On the other hand,
the discussion of velocity oscillations of neutrinos
presented in \cite{Glas} for the
low-energy region is compatible with our theory. However, the universality
of $c$ seems natural in unified
field theories (whereas that of the mass is
naturally violated) and preserves the Poincar\'e relativity principle
in the low-momentum limit. As previously stressed, deviations from 
the universality of $c$ due to nucleon or nucleus structure are expected to be
very small according to naive soliton models. 

For ultra-high energy
cosmic rays in the $E~\approx ~10^{20}~eV$ region, we expect the
most dramatic physical effects to be governed by the values of $\alpha $
for the particles considered. If Lorentz symmetry
is broken and an absolute rest frame exists, high-energy particles are
indeed different physical objects from low-energy particles, and 
high-energy tests of Lorentz symmetry are required. To reach direct 
comparison with a $\approx ~3.10^{20}~eV$ cosmic ray event, 
a $p~-~p$ collider
with energy $\approx ~400~TeV$ per beam would be required. Very-forward
experiments at LHC and VLHC 
would be crucial steps in a systematic check of the validity of Lorentz 
symmetry, comparing their data with those of cosmic-ray events above
$\approx ~10^{16}~eV$. Evidence for Lorentz symmetry violation would no
doubt be the most important physics outcome of particle physics experiments
in the decades to come. 

The possibility of taking the value of $a^{-1}$ close to the
wave vector of the highest-energy cosmic rays, i.e. 
$\approx 10^{26}~cm^{-1}$ , was considered in \cite{Gon9704} in connexion
with a possible search for DRK effects through particle lifetimes.
With  $\alpha ~\approx ~0.1$ and formulae (1)-(3), this would not be
incompatible with low-energy bounds on LSV. But the value of $E_{lim}$
would become too low leading to obvious incompatibilities with data in the
very high-energy region (the fall of final-state phase space). New bounds on
LSV thus emerge from high-energy data using the parametrization (1)-(3). 
Requiring that: a) cosmic rays with 
energies below $\approx ~3.10^{20}~eV$ deposit most of their energy in the 
atmosphere; b) the GZK cutoff is suppressed at energies above $\approx ~
10^{20}~eV$ , leads in the DRK scenario to the constraint: 
$10^{-72}~cm^2~<~\alpha ~a^2~<~
10^{-61}~cm^2$ , equivalent to $10^{-20}~<~\alpha ~<~10^{-9}$ for
$a~\approx 10^{-26}~cm$~.
Remarkably enough, assuming full-strength 
LSV forces $a$ to be in the range $10^{-36}~cm~<~a~<~
10^{-30}~cm$~. 
Data on high-energy cosmic
rays contain information relevant to the search for DRK signatures 
and should be carefully
analyzed. The energy dependence of the $\pi ^0$ lifetime above
$E~\approx ~10^{18}~eV$ can be a basic ingredient in generating the
specific cascade development profile (e.g.
electromagnetic showers versus hadronic showers and muons). 
Beyond DRK, strong signatures can be produced by other LSV effects: failure
of the parton model for protons and nuclei (including its dual 
"soft" version \cite{Aur1,Aur2}), substantially changing the 
multiplicities and event shape; strong deviations from the
relativistic formulae for Lorentz contraction and time dilation leading to 
basic modifications of the dynamics...
Because of its
stability at very high energy, the neutron becomes
a serious candidate to be a possible primary of the 
highest-energy cosmic-ray
events.

Cosmic rays seem to indeed be able
to test 
the predictions of (1)
and set upper bounds on the fundamental length $a$ , as well as constraints
on $\alpha $ .
Experiments like 
AUGER, OWL, AIRWATCH FROM SPACE 
and AMANDA present great potentialities.
The study of early cascade development (perhaps with balloons installed
in coincidence with the above experiments) will be
crucial for the proposed investigation.
Very high-energy data may even provide a way to measure
neutrino masses and mixing, as well as other parameters related to phenomena
beyond the standard model. If Lorentz symmetry were not violated, there would 
be no fundamental difference between the collision of a very high-energy 
cosmic ray and the "Lorentz equivalent" event at a collider. But, if Lorentz
symmetry is violated, the study of the parameters governing LSV will provide
us with a unique microscope directly focused on Planck-scale physics. Indeed,
the $E~\approx ~10^{20}~eV$ scale is closer, in order of magnitude,
to the Planck scale than to the electroweak scale. 

The possibility that above $E_{lim}$ cosmic rays
do not deposit all their energy in the atmosphere suggests 
to operate underground detectors in coincidence with air shower detectors,
if ever feasible at the required scales. At $E~\approx ~E_{lim}$ ,
the cosmic particle can still deposit enough energy
in the atmosphere to produce a detectable air shower for satellite-based 
and balloon experiments.

\end{document}